\documentclass[journal]{IEEEtran}
\usepackage{amssymb}
\usepackage{amsthm}

\usepackage{amsmath}
\usepackage{algorithm}
\usepackage{algorithmic}

\usepackage{bm}
\usepackage{graphicx}
\usepackage{float}
\usepackage{subfig}
\usepackage{mathrsfs}
\usepackage{diagbox}
\usepackage{mathtools}

 \usepackage{color}

 \renewcommand{\algorithmicrequire}{\textbf{Input:}}
  \renewcommand{\algorithmicensure}{\textbf{Output:}}

\usepackage{tikz}
\usepackage{multirow}
\usetikzlibrary{arrows}

\ifCLASSINFOpdf
\else
\fi
\begin{document}
\title{High-Performance Privacy-Preserving Matrix Completion for Trajectory Recovery}
\author{Jiahao Guo, An-Bao~Xu$^{*}$  

\thanks{This work was supported by the National Natural Science Foundation of China under Grant 11801418.}
\thanks{$^{*}$ Corresponding author}
\thanks{The authors are from college of Mathematics and Physics, WenzhouUniversity, Zhejiang 325035, China
(e-mail adress: guojiahao326@163.com (Jiahao Guo),
xuanbao@wzu.edu.cn (An-Bao Xu)).}}

%
%


\maketitle

\begin{abstract}
Matrix completion has important applications in trajectory recovery and mobile social networks. However, sending raw data containing personal, sensitive information to cloud computing nodes may lead to privacy exposure issue.The privacy-preserving matrix completion is a useful approach to perform matrix completion while preserving privacy. In this paper, we propose a high-performance method for privacy-preserving matrix completion. First,we use a lightweight encryption scheme to encrypt the raw data and then perform matrix completion using alternating direction method of multipliers (ADMM). Then,the complemented matrix is decrypted and compared with the original matrix to calculate the error. This method has faster speed with higher accuracy. The results of numerical experiments reveal that the proposed method is faster than other algorithms.
\end{abstract}

\begin{IEEEkeywords}
Privacy-preserving matrix completion,
alternating direction method of multipliers, trajectory recovery \end{IEEEkeywords}

\IEEEpeerreviewmaketitle

\section{INTRODUCTION}
With the advancement of information technology, there is a growing demand for enhanced analysis and processing capabilities in handling large-scale data. Various fields such as system identification \cite{ZMM2012}, image processing \cite{WSL2022,WSC2022}, and recommendation systems \cite{ZXL2022} regularly encounter extensive datasets containing valuable information awaiting extraction. However, the processing of large-scale data often faces challenges such as missing elements, contamination, and data corruption during acquisition and transmission \cite{KXL2013}. Effectively addressing these issues to accurately and efficiently recover and process incomplete or corrupted data holds significant practical importance.
Matrix completion \cite{C2015,CR2012,KS2023} is able to fill the matrix containing a large number of missing elements almost perfectly. The matrix completion problem usually involves a large amount of computation, and in computing it is common to outsource such a large computational task to a cloud server. In cloud computing, a user sends an ordinary data request from a mobile device to a cloud node, which performs a matrix completion algorithm and finally sends it from the cloud node to the user, potentially exposing the user's sensitive information to the cloud node. However, if a user sends data to the server or publishes anonymised data, this raises privacy concerns \cite{B2018}. In this case, the user's privacy cannot be well protected and there is a risk of leakage \cite{XZ2022,WSR2022}.

Privacy-Preserving computation \cite{LLJ2016}, refers to the use of cryptography \cite{B2012,CKKS2017}, trusted hardware \cite{ZWW2021,PS2019}, multi-party secure computing \cite{CGR2017,Y1982}, differential privacy \cite{JGS2021,DR2014} and other cross-fertilisation techniques to analyse and compute the privacy of data under the premise of ensuring data privacy and security. Computing technology does not refer to a particular technology, but to a system of technologies in general. Through the use of many cross-fertilisation techniques, privacy computing can realise the availability of data without visibility, and achieve the purpose of data security flow to realise the value of data. Therefore, based on the principle of security mechanism, all the key technologies are classified. They are mainly divided into two major fields: cryptography and trusted hardware. Cryptography technology is represented by multi-party secure computing; trusted hardware mainly refers to trusted execution environment. In addition, there are also related application technologies such as federated learning \cite{ZMH2023,LY2022} derived from the above two technology paths.

Currently, matrix completion and privacy-preserving computation are maturely developed. Liu $et~al. ~$\cite{LL2022} proposed a homomorphic encryption-decryption scheme which is theoretically proved that the scheme satisfies the homomorphic and differential privacy properties. Besides, a GPU-based homomorphic matrix completion scheme to achieve high precision and accuracy with guaranteed privacy has been designed by Zhang $et~al.$ \cite{ZLL2020}. Kong $et~al. ~$\cite{KHL2015} first proposed a homomorphic method to encrypted the incomplete matrices and then complete the processed matrices by using the alternating minimization algorithm (ALT-MIN)~\cite{JNS2013,BT2013}. Using this approach, it is not only possible to reconstruct incomplete data from different users on cloud computing nodes, but also to protect user's privacy at the same time. Currently, a commonly used privacy-preserving technology is anonymisation, which protects user privacy to a certain extent, but in order to further improve privacy, we add fake data and perturb the original data; and from \cite{LDY2011}, we have known that $L_{2,1}$-norm have higher accuracy in solving matrix completion problems. Based on the above propose, this paper proposes a new privacy-preserving matrix completion method based on matrix tri-factorization to solve the $L_{2,1}$-norm minimization problem.

Our contributions are as follows:
\subsubsection{}

In order to complete the matrix completion quickly and protect the user's privacy at the same time, a privacy-preserving method based on CSVD-QR-based $L_{2,1}$-norm minimisation (PPLNM-QR) is proposed. By using QR as an alternative to singular value decomposition (SVD), PPLNM-QR is much faster and more accurate than traditional methods using SVD.
\subsubsection{}
We introduce a novel framework for privacy-preserving matrix completion (PPMC), which integrates an encryption methods with matrix completion techniques to achieve both robust privacy preservation and high accuracy in results. This framework is designed to minimize computational overhead while ensuring efficient and precise outcomes, making it accessible for implementation across various applications.

In the next section, we shall introduce some notations and preliminaries. We present our algorithm and framework in Section III. Section IV describes the experimental results. We finally get our conclusion in Section V.

\section{NOTATIONS AND PRELIMINARIES}
\subsection{{Notations}}
In this paper, we denote matrices by boldface capital letters (e.g.,$\ M\in{\mathbb{R}}^{S \times T}$) and vectors by boldface lowercase letters (e.g.,$ m\in{\mathbb{R}}^{S}$). We use $~\tilde{M}$ to indicate that this is the encrypted matrix. We use $~\hat{{M}}$ to indicate that this is the decrypted matrix. [$K$] denotes the set\{$1,2,\ldots,k$\}.we use $i,j$ to index the rows and columns of a matrix. For example,a matrix $ M\in{\mathbb{R}}^{S \times T}$, $M_{i}$ or $ M(S,:)$ denotes the $i$-th row of $ M$, the $(i,j)$ -th element is $ M(i,j)$ or $M_{ij}$, where$i\in[S]$ and $j\in[T]$. We use $M^T$ to denote the transposed matrix of $M$. The Frobenius norm of a matrix $ M$ is $|| M||_F=\sqrt{\sum_{i=1}^{S}\sum_{j=1}^{T}|M_{ij}|^2}$. The $L_{2,1}$-norm of the matrix $M$ is $||M||_{2,1}=\sum^{S}_{i=1} \sqrt{\sum_{j=1}^{T}M_{ij}^2}=\sum^{S}_{i=1}|| M_{i,:}||_2$. We use $ M^i$ to denote the $i$-th matrix in a series of matrices.

\subsection{{Matrix Completion Model}}
Matrix completion was first proposed by Candššs $et~al.$in 2008 \cite{CR2012}, assuming that the matrix $M\in{\mathbb{R}}^{S \times T}$ has the following form:

\begin{figure}[htbp]
\centering
\includegraphics[width=140pt]{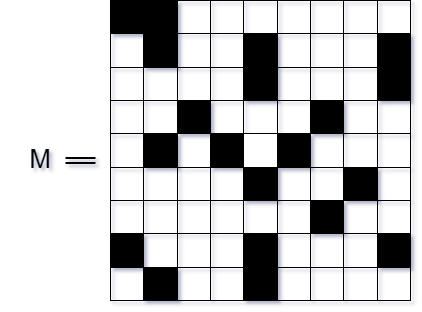}
\end{figure}

\noindent
where black denotes the known elements of the matrix and white denotes the unknown elements of the matrix known elements of the matrix. If the matrix $M$ has the property of low rank, the problem can be transformed into solving a convex optimisation problem. Matrix completion is an affine rank minimisation problem under random constraints, and the general affine rank minimisation problem is described as follows: \begin{align}  \mathop{\min}\limits_{X}~~~~~ \rm rank(X) ~~~~s.t. ~~~ X_{i,j}= M_{i,j},{i,j} \in \Omega \end{align}
where $\Omega$ is the set of locations corresponding to the observed entries.

\subsection{{Index Matrix}}
An index matrix is a matrix which denotes whether an element $X_{ij}\in X $ is missing,which is defined as:
\begin{equation}
\Omega_{ij}=\begin{cases} 
1,&\ otherwise\\
0,&\ if ~~ X_{ij} ~~ is ~~ missing
\end{cases}
\end{equation}

\subsection{{Hadamard Product}}
The Hadamard product is a class of operations on matrices that operate on matrices of the same shape and produce a third matrix of the same dimension. An example of this product of two matrices $ X$ and $\Omega$ of size ${S}\times{T}$ is defined as:
$$  X\circ\Omega= \begin{bmatrix}
      X_{11}\Omega_{11}  & \cdots &  X_{1{T}}\Omega_{1{T}} \\
    \vdots  &    \ddots & \vdots \\
      X_{{S}1}\Omega_{{S}1}  & \cdots  & X_{{S}{T}}\Omega_{{S}{T}} \\
\end{bmatrix}  \in{\mathbb{R}}^{S \times T},$$



\section{FRAMEWORK OF PRIVACY-PRESERVING MATRIX COMPLETION FRAMEWORK}
The privacy-preserving matrix completion framework mainly consists of two parts, one is on the user mobile devices, the other is on the cloud computing nodes. The details of the privacy-preserving matrix completion framework on user devices and on cloud computing nodes are shown in algorithm $1$ and $2$.

\begin{figure}[htbp]
\centering
\includegraphics[width=280pt]{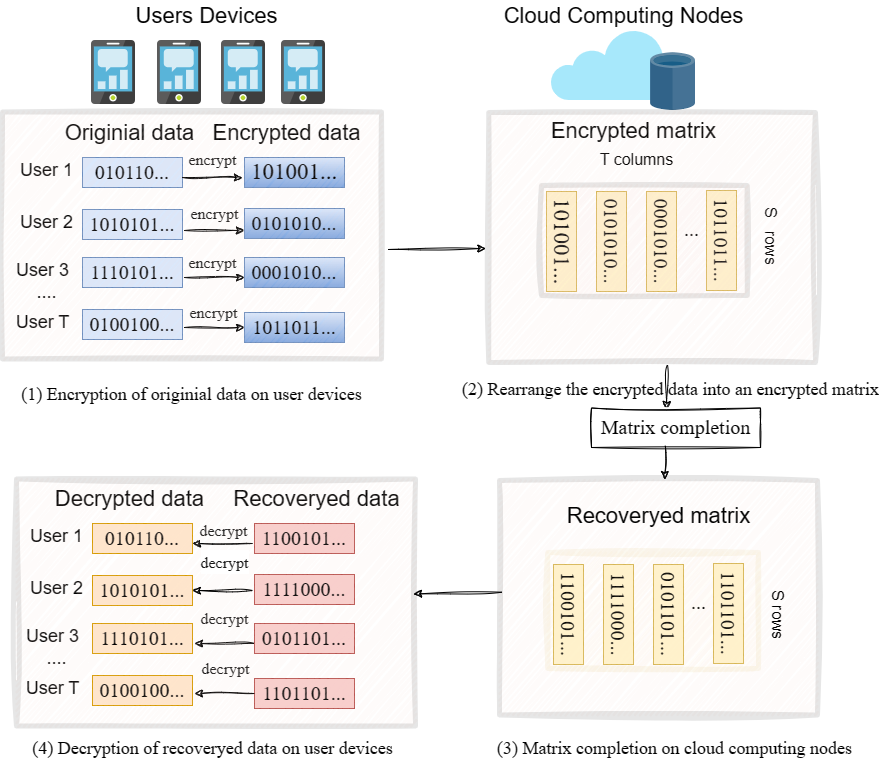}
\caption{Framework of privacy-preserving matrix completion framework}
\end{figure}

The operations on the mobile device is shown in fig.$1$:

First, user receives the public matrix $ P\in\mathbb{R}^{S \times T}$from cloud computing nodes and encrypts the original data with $P$ on his own device, then the user sends the encrypted data to the cloud. Each user's data can be viewed as an encrypted vector and the cloud computing nodes reconstruct these vectors into an encrypted matrix after accepting data from all users. Second, the algorithms on the cloud computing nodes are utilised to perform matrix completion on the reconstituted encrypted matrix to recover the encrypted matrix with the goal of low rank R. Third,users download the recovered and encrypted matrix$\textbf~\hat{M}$ from the cloud nodes. Users decrypt the columns of the downloaded matrix on the devices using their own private keys to obtain their own original data.

The first and third steps use privacy-preserving encryption and decryption to encrypt the original data and decrypt the reconstituted matrix to protect the privacy of the data. In the second step, matrix completion is performed by using ADMM \cite{HTY2012,LLS2013}.

\begin{algorithm}[!h]
    \caption{Privacy-Preserving Encryption and Decryption Algorithm on User $k$, for $k\in[K]$}
    \label{alg:AOA}
    \renewcommand{\algorithmicrequire}{\textbf{Input:}}
    \renewcommand{\algorithmicensure}{\textbf{Output:}}
    \begin{algorithmic}[1]
        \REQUIRE The original vector $ M_k\in{\mathbb{R}}^{S}$and index vector $\Omega_t\in{\mathbb{R}}^{S}$. 
        \ENSURE The recovered and complete data vector$~\hat{D_k}$.   
        \STATE Initialize:Receive the public keys $P\in{\mathbb{R}}^{S \times T}$ from the cloud computing nodes;
        \STATE Randomly generate $\psi_0,\psi_1,\cdots,\psi_I\in[0,1]$
        as his private keys and
        $\sum^{I}_{i=0}\psi_i=1$;
        \STATE With the public matrix and private keys,encrypt the $ M_k$ to $\textbf~\tilde{M}_k$ by $\textbf~\tilde{M}_k=(\psi_0 M_k+\psi_1P_1+\cdots+\psi_I P_I)\circ\Omega_t$;
        \STATE Upload encrypted data $\textbf~\tilde{M}_k$ to the cloud computing nodes;
        \STATE Cloud nodes reconstruct each encrypted data vector into an encryption matrix$\textbf~\tilde{M}\in\mathbb{R}^{S \times T}$;
        \STATE Download the recovered and encrypted matrix $\textbf~\hat{M}$ from the cloud computing nodes;
        \STATE Decrypt $\textbf~\hat{M_k}$ to $~\hat{D}_k$ by
        $\textbf~\hat{D}_k=(\textbf~\hat{M_k}-\psi_1 P_1-\cdots-\psi_IP_I)/\psi_0$.
    \end{algorithmic}
\end{algorithm}

\subsection{Encryption of Original Data}~
Details of encryption of original data on user devices is shown in lines $1$-$3$ of algorithm $1$. For example, a user $k$ randomly receives his public matrix $P\in{\mathbb{R}}^{S \times I}$ from cloud computing nodes, $I$ denotes the rank of the public matrix $P$. Split the matrix $P$ into $I$ columns $ P_1, P_2, P_3,\cdots, P_I$ as his public keys. Then, user $k$ generates $I+1$ private keys $\psi_0,\psi_1,\cdots,\psi_I$ on his own devices. And the private keys satisfies $\psi_i\in[0,1]$ and $\sum^{I}_{i=0}\psi_i=1$. Based on these keys, user $k$ encrypts his original data vector $M_k\in{\mathbb{R}}^{S}$ to obtain an encrypted data vector $\textbf~\tilde{M}_k\in{\mathbb{R}}^{S}$. The encryption formula is shown below:
 \begin{equation}\label{14}
 \textbf~\tilde{M}_k=(\psi_0 M_k+\psi_1 P_1+\cdots+\psi_IP_I)\circ\Omega_t\in{\mathbb{R}}^{S}
 \end{equation}
This privacy-preserving encryption scheme which called light-weight encryption scheme does not need too much communication overhead¡¢computing power¡¢ memory capacity and power budget. The above steps are the encryption operations performed by the user at his own device. Then, the $K$ users upload their encrypted data vectors$\textbf~\tilde{M}_k\in{\mathbb{R}}^{S}$to cloud computing nodes and the nodes rearrange the encrypted data vectors into an encrypted matrix $\textbf~\tilde{M}$, where each encrypted data vector becomes a column of the encrypted matrix $\textbf~\tilde{M}$.

\begin{algorithm}[!h]
    \caption{Privacy-Preserving $L_{2,1}$-norm Minimisation Method on Cloud Computing Nodes (PPLNM-QR)}
    \label{alg:AOA}
    \renewcommand{\algorithmicrequire}{\textbf{Input:}}
    \renewcommand{\algorithmicensure}{\textbf{Output:}}
    \begin{algorithmic}[1]
        \REQUIRE  Encrypted matrix $\textbf~\tilde{M}\in\mathbb{R}^{S \times T}$;
        Index vector $\Phi_t\in\mathbb{R}^{S\times T}$.
        \ENSURE The recovered matrix$\textbf~\hat{M}$    
        \STATE {\textbf{Initialize}}:~k=0;~Itmax\textgreater 0; ~q\textgreater 0;~$\mu\textgreater 0$;~$ \rho \textgreater 0$;
        R(The rank of the matrix$\textbf~\tilde{M}\in\mathbb{R}^{S\times T}$);
        $\epsilon$ is a positive tolerance;
        $~\tilde Y\in\mathbb{R}^{S\times T}$;
        $~\tilde L_1=\rm eye(S,R)$;
        $~\tilde D_1=\rm eye(R,R)$;
        $~\tilde R_1=\rm eye(R,T)$;
        \WHILE {${||~\tilde X_k-~\tilde X_{k+1}||_F^2} \geq \epsilon$ and $k \textless l$}
                \STATE $[~\tilde L_{k+1},\sim]=qr((~\tilde X_k+\frac{~\tilde Y_k}{u_k})*R_k^T)$;
                \STATE $[~\tilde R_{k+1},~\tilde D_{T}^T]=qr((~\tilde X_k+{~\tilde Y_k})^T*L_k)$;
                \STATE $~\tilde R_{k+1}=~\tilde R_{k+1}^T$;
        \FOR {$k=1$ to $K$}
                \FOR {$j=1$ to $R$}
                \STATE $~\tilde D_{j+1}=\frac {(||{~\tilde D_T}^j||_F-\frac {1}{u_k})}{||{~\tilde D_T}^j||_F}$;
                \ENDFOR
        \ENDFOR
              \STATE $~\tilde X_{k+1}=~\tilde L_{k+1}*~\tilde D_{k+1}*~\tilde R_{k+1}-\Omega_t*(~\tilde L_{k+1}*~\tilde D_{k+1}*~\tilde R_{k+1})+~\tilde{M}$;
              \STATE $~\tilde Y_{k+1}=~\tilde Y_k+\mu_k(~\tilde X_{k+1}-~\tilde L_{k+1}*~\tilde D_{k+1}*~\tilde R_{k+1})$;
              \STATE $\mu_{k+1}=\rho\mu_k$;
     \ENDWHILE
        \RETURN $~\hat{M}=~\tilde X_{k}$;
        \STATE Send the$\textbf~\hat{M_k}$ to the corresponding user.
    \end{algorithmic}
\end{algorithm}

\subsection{Privacy-Preserving Matrix Completion}~
The algorithm for $L_{2,1}$-norm minimisation method is shown in algorithm $2$. After the cloud computing node reorganises the encrypted vectors into an encrypted matrix $\textbf~\tilde{M}$ , we need to complete this incomplete matrix. From \cite{LLY2010}, we know the $L_{2,1}$-norm was successfully used in low rank representation to optimize the noise data matrix $\textbf~\tilde{M}$ .The optimal $\textbf~\tilde{M}$ can be updated by solving the minimization problem as follows:\begin{align}  \mathop{\min}\limits_{~\tilde{M}}~ \frac{1}{\mu} ||~\tilde{M}||_{2,1}+\frac {1}{2}||~\tilde{M}-Y||_F^2\end{align}
where $\textbf{Y}\in{\mathbb{R}}^{S \times T}$ is a given real complete matrix and $\mu \textgreater 0$.
First, we introduce a matrix decomposition method PPCSVD-QR. Suppose that $\textbf~\tilde {M}\in{\mathbb{R}}^{S \times T}$ is a given real matrix, we can find three matrices, i.e., $\textbf ~\tilde L,\textbf ~\tilde D,\textbf ~\tilde R$ such that\begin{align}\textbf ~\tilde M=\textbf ~\tilde L*\textbf ~\tilde D*\textbf ~\tilde R\end{align}
where $\textbf ~\tilde L\in R^{S\times R}$ is a column orthogonal matrix, $\textbf ~\tilde D\in \mathbb{R}^{R\times R}$ is a diagonal matrix and $\textbf ~\tilde R\in \mathbb{R}^{R\times T}$ is a row orthogonal matrix.
Suppose that the results of the $k$ -th iteration are $ \textbf ~\tilde L_{k}, \textbf ~\tilde D_{k}, $ and $\textbf ~\tilde R_{k} $ with $\textbf ~\tilde L_{0}=\rm eye(S,R)$, $\textbf ~\tilde D_{0}=\rm eye(R,R)$, and $\textbf~\tilde R_{0}=\rm eye(R,T)$. First, $~\tilde L_{k}$ is given by QR decomposition of $\textbf ~\tilde {M*R}_{k-1}^{T}$ as
\begin{align} \textbf ~\tilde {M} * ~\tilde{R}_{k-1}^{T}=\textbf ~\tilde L_{k}*\textbf ~\tilde D_k \end{align}
Then,$\textbf ~\tilde R_{k+1}$ is given by the same way
\begin{align}\textbf ~\tilde M^{T}*\textbf ~\tilde L_{k}=\textbf ~\tilde R_{k}*\textbf ~\tilde D_{k} \end{align}
Finally, $\textbf  ~\tilde D_{k}$ is updated by
\begin{align}\textbf ~\tilde D_{k}=\textbf ~\tilde D_{k}^{T},  \end{align}
Then, we use the obtained tri-factorization matrices to rewrite our equation (4) :
\begin{align} &{\mathop{\min}\limits_{~\tilde{L},~\tilde{D},~\tilde{R}}~ \frac{1}{\mu} ||\textbf ~\tilde L*\textbf ~\tilde D*\textbf ~\tilde R||_{2,1}+\frac {1}{2}||\textbf ~\tilde L*\textbf ~\tilde D*\textbf ~\tilde R-\textbf ~\tilde Y||_F^2} \nonumber\\
& {s.t. ~~\textbf ~\tilde M=\textbf ~\tilde L*\textbf ~\tilde D*\textbf ~\tilde R}
\end{align}
Since both $\textbf ~\tilde L$ and $\textbf ~\tilde R$ are orthogonal, the problem can again be rewritten as:
\begin{align} &{\mathop{\min}\limits_{~\tilde{D}}~ \frac{1}{\mu} ||\textbf ~\tilde D||_{2,1}+\frac {1}{2}||\textbf~\tilde D-\textbf ~\tilde E||_F^2} \nonumber\\
&{s.t. ~~\textbf ~\tilde M=\textbf ~\tilde L*\textbf ~\tilde D*\textbf ~\tilde R ~~, ~~\textbf ~\tilde E=\textbf ~\tilde L^T* ~\tilde Y *\textbf~\tilde R^T} \end{align}
For problem (10), it has the following expression for the contraction operator: \begin{align}
\tilde D(:,j)=\frac {(||~\tilde E(:,j)||_F-\frac{1}{\mu})_+}{||~\tilde E(:,j)||_F} ~\tilde E(:,j)\end{align}
where\begin{align}(x)_+= \mathop{\max}\{x,0\}
\end{align}with $x\in(-\infty,+\infty)$ being a real number.

By fixing the variables$~\tilde L_{k+1},~\tilde D_{k+1},~\tilde R_{k+1}$, we can update the $~\tilde X_{k+1}$ as follows:
\begin{align}
&{~\tilde X_{k+1}=~\tilde L_{k+1}*~\tilde D_{k+1}*~\tilde R_{k+1}}\nonumber \\
&{-\Omega_t*(~\tilde L_{k+1}*~\tilde D_{k+1}*~\tilde R_{k+1})+ ~\tilde{M}}
\end{align}
Then by fixing variables $~\tilde Y_{k},~\mu_k$, we update $~\tilde Y_{k+1}$ and ${\mu_{k+1}}$ as follows:
\begin{align}
~\tilde Y_{k+1}=~\tilde Y_k+\mu_k(~\tilde X_{k+1}-~\tilde L_{k+1}*~\tilde D_{k+1}*~\tilde R_{k+1})
\end{align}
\begin{align}
\mu_{k+1}=\rho\mu_k
\end{align}
where $\rho\geq 1$. PPLNM-QR mainly uses ADMM to solve the convex optimisation function in (10) which is a gradient search based method and hence the method converges to its optimal solution.

Finally, the cloud computing nodes send the recovered vector $\textbf~\tilde{M_k}$ to the $k$-th user, for $k\in[K]$.

\subsection{ Decryption of Reconstructed Data}~
Details of decryption of encrypted data on user devices is shown in lines $6$-$7$ of algorithm $1$. When the encrypted data is recovered on the cloud computing nodes, the corresponding user can download the corresponding encrypted data and apply the decryption operation. For example, the user $k$ utilizes his private keys $\psi_0,\psi_1,\cdots,\psi_I$ and public keys $ P_1, P_2, P_3,\cdots, P_I$ to decrypt$\textbf~\hat{M_k}$ as formula showing below:
 \begin{equation}\label{14}
 \textbf~\hat{D}_k=(\textbf~\hat{M_k}-\psi_1 P_1-\cdots-\psi_I P_I)/\psi_0\in{R}^{S}
 \end{equation}
The$\textbf~\hat{D}_k$ which is similar to$\textbf~{M}_k$ (the complete and original data of user) is only known by user $k$.

\section{EXPERIMENTAL RESULTS}
To demonstrate the feasibility of our method, we performed several comparative experiments on synthetic and real datasets to compare the results with run times and recovery errors (RSE) ,i.e.,$\rm RSE=\frac{\|M-\textbf~\hat {D}\|_F}{\|D\|_F}$,where $~\hat{D}, M\in{\mathbb{R}}^{S \times T}$ are the recoveryed matrix and real matrix respectively.

The experiments are all run on a MATLAB 2019a platform equipped with an i5-12500 CPU and 16GB of RAM.

\subsection{{Experiments with a Synthetic Dataset}}~
We randomly generate matrices of different sizes ($S\times T$) to evaluate the performance of our algorithm, which the matrix satisfies the property of approximately low rank. So it is feasible to use a matrix completion approach to solve this problem efficiently. The rank of matrices $R$ is set to be $0.01\times min(S,T)$ and the data loss rate is fixed at $0.5$. The speed of the algorithm is expressed by the interval of one iteration of an algorithm, the recovery accuracy of the algorithm is expressed by RSE.

For convenience, we fixed the matrix size to compare the running time of the two algorithms as well as the RSE obtained, with matrix sizes ranging from $128$ to $8192$.
Table \uppercase\expandafter{\romannumeral1} shows the running time and RSE obtained by the two algorithms after different iterations ($K$=$100$-$300$).

After performing the calculations, we found that the difference between the running time of the two algorithms is huge, and in order to make the image aesthetically pleasing, we do logarithmic processing of the running time. Fig.$2$ depicts the running time of the two algorithms for numerous iterations.

Obviously, the current method is much faster than the other methods, with a huge increase in speed,while the precision is also satisfactory. This result is the same as expected, mainly because the complexity of the two algorithms is different. Let's briefly analyse the computational complexity of ALT-MIN and PPLNM-QR. The computational cost of SVD on$~\tilde M$ is $O(ST^2)$. The main time of ALT-MIN is consumed by performing SVD on$~\tilde M$. So, its computational complexity is $O(ST^2)$. The computational cost of QR on $~\tilde M$ is $O(r^2(S+T))$, which $r \ll min(S,T)$. The iterative process of PPLNM-QR is performing two QR decompositions on $~\tilde M$ to update $~\tilde L$ and $~\tilde R$ ((6)-(8)). Thus, the computational complexity of PPLNM-QR is $O(r^2(S+T))$. Obviously, the computational complexity of PPLNM-QR is much smaller than that of ALT-MIN. Therefore, PPLNM-QR will be much faster than those SVD-based algorithms (ALT-MIN,Singular Value Thresholding (SVT)).

\begin{figure}[htbp]
\centering
\includegraphics[width=215pt]{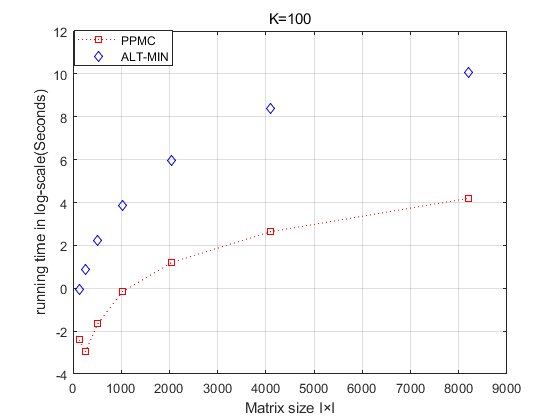}
\includegraphics[width=215pt]{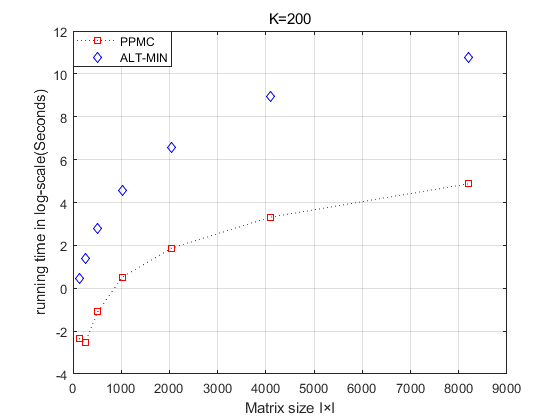}
\includegraphics[width=215pt]{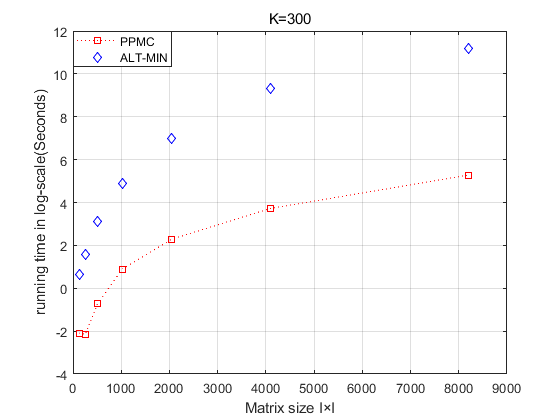}
\caption{Running time for the synthetic dataset}
\end{figure}

\begin{table}[h!]
\centering
\caption{RUNNING TIME AND RSE FOR THE SYNTHETIC DATASET }
\scalebox{1.0}{
\label{table1}
\begin{tabular}{|c|c|c|c|c|}
\hline
{}& \multicolumn{4}{c|}{Number of Iterations (K=100)}\\

\cline{2-5}

{} & \multicolumn{2}{c|}{ALT-MIN}& \multicolumn{2}{c|}{PPMC}  \\
\hline
{MATRIX SIZE} & TIME & RSE & TIME & RSE  \\
\hline

$128\times128$ &0.9447 & 8.8985e-16&	0.0895 &	4.2670e-15  \\
\cline{1-5}

$256\times256$&2.3731	&1.2644e-15	&0.0512	&7.9315e-16\\
\cline{1-5}

$512\times512$&9.1006	&1.9172e-15	&0.1899	&8.0248e-16\\
\cline{1-5}

$1024\times1024$&48.2394	&1.7696e-15	&0.8478	&1.0557e-15\\
 \cline{1-5}

$2048\times2048$&389.9873	&1.5858e-15	&3.3307	&1.1312e-15\\
 \cline{1-5}

$4096\times4096$&4336.1026	&1.9088e-15	&13.9924	&1.1570e-15\\
 \cline{1-5}

$8192\times8192$&23965.8257	&2.1934e-15	&66.0265	&1.5134e-15\\
 \cline{1-5}
\hline
\end{tabular}
\label{table_MAP}
}
\end{table}

\begin{table}[h!]
\centering

\scalebox{1.0}{
\label{table1}
\begin{tabular}{|c|c|c|c|c|}
\hline
{}& \multicolumn{4}{c|}{Number of Iterations (K=200)}\\

\cline{2-5}

{} & \multicolumn{2}{c|}{ALT-MIN}& \multicolumn{2}{c|}{PPMC}  \\
\hline
{MATRIX SIZE} & TIME & RSE & TIME & RSE  \\
\hline

$128\times128$ &1.5723 & 7.9325e-16 &	0.0956 &	5.6453e-16  \\
\cline{1-5}

$256\times256$&3.9460	&1.2369e-15	&0.0812	&1.1037e-15\\
\cline{1-5}

$512\times512$&16.5783	&1.9070e-15	&0.3318	&7.6596e-16\\
\cline{1-5}

$1024\times1024$&93.5302	&1.7440e-15	&1.6885	&1.0023e-15\\
 \cline{1-5}

$2048\times2048$&724.4106	&1.5793e-15	&6.5180	&9.5880e-16\\
 \cline{1-5}

$4096\times4096$&7585.0207&1.9094e-15&27.6081&9.8635e-16\\
 \cline{1-5}

$8192\times8192$&47793.7349&2.1861e-15&130.2746&1.4275e-15\\
 \cline{1-5}
\hline
\end{tabular}
\label{table_MAP}
}
\end{table}

\begin{table}[h!]
\centering
\scalebox{1.0}{
\label{table1}
\begin{tabular}{|c|c|c|c|c|}
\hline
{}& \multicolumn{4}{c|}{Number of Iterations (K=300)}\\

\cline{2-5}

{} & \multicolumn{2}{c|}{ALT-MIN}& \multicolumn{2}{c|}{PPMC}  \\
\hline
{MATRIX SIZE} & TIME(S) & RSE & TIME(S) & RSE  \\
\hline

$128\times128$ &1.8720 & 8.6589e-16 &	0.1238 &	 4.6956e-16  \\
\cline{1-5}

$256\times256$&4.8448	&1.2433e-15	&0.1171	&7.5387e-16\\
\cline{1-5}

$512\times512$&22.8095	&1.9050e-15	&0.4905	& 8.7933e-16\\
\cline{1-5}

$1024\times1024$&130.8889	&1.7489e-15	&2.4546	& 8.1087e-16\\
 \cline{1-5}

$2048\times2048$&1076.9538	&1.6135e-15	&9.8745	&8.7646e-16\\
 \cline{1-5}

$4096\times4096$& 11129.0325&1.8864e-15&41.4339&1.0690e-15\\
 \cline{1-5}

$8192\times8192$&70585.7261 &2.2042e-15&195.5440&1.4673e-15\\
 \cline{1-5}
\hline
\end{tabular}
\label{table_MAP}
}
\end{table}

\subsection{{Experiments with a Real Dataset}}~

\textbf{Real Image Completion.} We randomly select a number of images to evaluate the effectiveness of our algorithm. The number of iteration steps $K$ are fixed at $300$ and the data loss rate $\alpha$ is fixed at $0.5$. Table \uppercase\expandafter{\romannumeral2} shows the running time and RSE. 

\begin{table}[h!]
\centering
\caption{RUNNING TIME AND RSE FOR THE REAL DATASET }
\scalebox{1.0}{
\label{table1}
\begin{tabular}{|c|c|c|}
\hline
{}& \multicolumn{2}{c|}{Number of Iterations (K=300)}\\

\cline{2-3}

{} &  \multicolumn{2}{c|}{PPMC}  \\
\hline
{VALUE} & TIME & RSE  \\
\hline

$321\times481\times3$ &	0.241668 &5.5616e-16  \\
\cline{1-3}
$512\times512\times3 $&0.397930	& 9.6663e-16	\\
\cline{1-3}
$1800\times1200\times3$&4.705361	&8.1990e-16	\\
 \cline{1-3}
$3000\times3000\times3$&20.950438	& 9.3366e-16	\\
 \cline{1-3}
$3145\times2489\times3$	&23.812629	&9.7043e-16\\
\cline{1-3}
$5088\times3168\times3$	&40.446035	&1.0357e-15\\
\cline{1-3}
\hline
\end{tabular}
\label{table_MAP}
}
\end{table}

\textbf{Traffic Data Completion.} We evaluate our algorithm based on a real-world traces dataset in Geolife\cite{GeoLife}. We pre-process the raw Geolife data for our recovery by selecting complete trajectories as our ground truth. The trajectories of $50$ users were selected for the test dataset and the length of time intercepted was $1175$ seconds with a time interval of 5 seconds for one node. The data loss rate $\alpha$ is fixed at $0.5$, the public trajectory is set to be $5$, the RSE was used as a quantitative metric.

In order to clearly show the results obtained with PPMC, we randomly selected the trajectory of one user (whose ID is $000$), ~fig.3 (a) shows the user's original trajectory, (b) shows the trajectory with four missing points,(c) shows the encrypted and missing trajectory and (d) shows the comparison between the recovered trajectory and the original trajectory using PPMC, using a $9$-location original trajectory. (Originally, there were 10 positions, two of which largely overlapped, so we manually eliminated one position, so only nine trajectories are shown.) We can see that when $50$ percent of the original data is missing, PPMC still recovers the original trajectories with high accuracy, as shown in 3(d). Comparing the latitude and longitude in 3(a) and 3(c), we find that the resulting trajectories are different and the locations are different, which shows that the encryption scheme can protect the user's privacy well.

To gain more insight into the effect of data loss on the recovery accuracy, we applied PPMC on the trajectory where $\alpha$ varies from $0.1$ to $0.9$, and the results are shown in Fig.$4$. As the value of $\alpha$ increases, there is a clear tendency for the recovery error to increase.

In summary, PPMC can not only achieve trajectory recovery quickly and accurately, but also can form a good protection for users' personal data.

\begin{figure}[!t]
\centering
\subfloat[The original trajectory]{
		\includegraphics[scale=0.27]{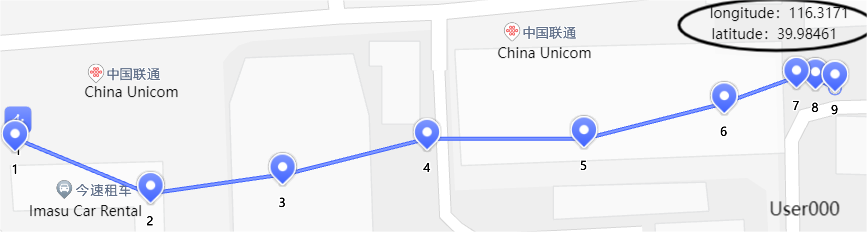}}
\\
\subfloat[The missing trajectory]{
		\includegraphics[scale=0.3]{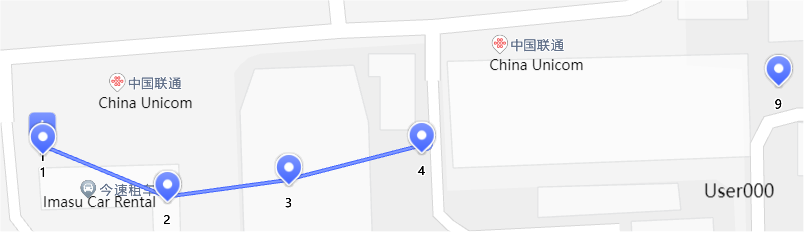}}
\\
\subfloat[The encrypted and missing trajectory]{
		\includegraphics[scale=0.45]{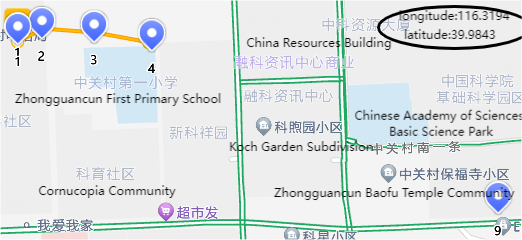}}
\\
\subfloat[Comparison of original trajectory and recovered trajectory]{
		\includegraphics[scale=0.2]{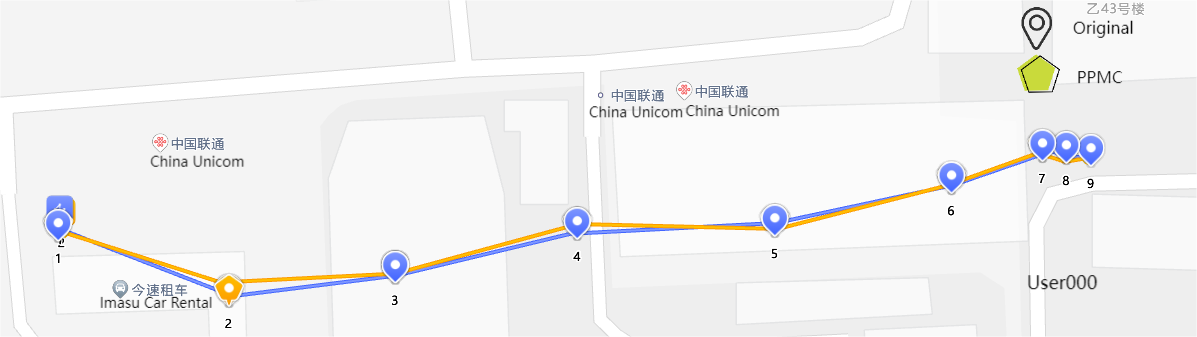}}
\caption{ Results showcase (The lines are the map matching results based on the dots) }
\label{fig_5}
\end{figure}

\begin{figure}[htbp]
\centering
\includegraphics[width=215pt]{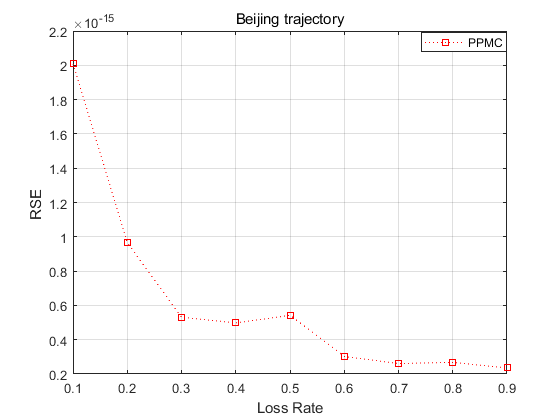}
\caption{Recovery error vs data loss rate}
\end{figure}

\section{CONCLUSION}
In this paper, we propose a fast and accurate framework for privacy preserving matrix completion. To improve the matrix speed, we use CSVD-QR instead of SVD, which is much faster than the traditional SVD. To improve the accuracy, we have known that $L_{2,1}$-norm have higher accuracy in solving matrix completion problems. In order to protect the user's privacy, we apply privacy-preserving techniques to matrix completion. Based on these properties, PPLNM-QR is proposed, which has several key advantages over ALT-MIN, including significantly faster processing time and smaller accuracy loss. In addition, we evaluate the method using synthetic and real datasets. The computational results show that the method is fast and accurate. While we focus on trajectory  recovery in this work, PPMC can also be used for other data recovery applications that maintain privacy protection,such as recommendation systems as well as image recovery.



 \newpage

\newpage

\newpage


 \newpage

\newpage

\label{5.2}



\end{document}